\documentclass[showpacs,preprint,aps]{revtex4}
\usepackage{graphicx}
\usepackage{dcolumn}
\usepackage{bm}
\begin{document}
\setcounter{page}{1}
\title 
{Quantum time scales in alpha tunneling}
\author
{N. G. Kelkar$^1$, H. M. Casta\~neda$^{1,2}$ and M. Nowakowski$^1$}
\affiliation{$^1$ 
Departamento de Fisica, Universidad de los Andes, 
Cra.1E No.18A-10, Bogota, Colombia\\
$^2$ Max-Planck-Institut f\"ur Kernphysik, 
Saupfercheckweg 1, 69117 Heidelberg, Germany
}
\begin{abstract}
The theoretical treatment of alpha decay by Gamow 
is revisited by investigating the quantum time scales in tunneling. 
The time spent by an alpha particle in front of the barrier 
and traversing it before escape is evaluated using microscopic 
alpha nucleus potentials. The half-life of a nucleus is shown to 
correspond to the time spent by the alpha knocking in front of the 
barrier. 
Calculations for medium and super heavy nuclei show that 
from a multitude of available tunneling time definitions,  
the transmission dwell time 
gives the bulk of the lifetime of the decaying state, in most cases.
\end{abstract}
\maketitle 

The concept of quantum tunneling has been with us for decades and has 
been successfully applied in many branches of physics. It essentially 
expresses the fact that in quantum mechanics there is always a non-zero 
finite probability for a particle to go from one region to another even if
the regions are separated by a large but finite potential barrier and the 
particle carries a kinetic energy less than the height of the barrier. Some 
of the very first applications were by Gamow \cite{gamow} and by Gurney and 
Condon \cite{gurney} 
to study the alpha decay of radioactive nuclei. It was 
however soon noticed that the tunneling phenomenon was not restricted to 
nuclear physics, but was rather a general result of quantum mechanics 
which is now commonly used to study the physics of semiconductors and 
superconductors, in constructing electron tunneling microscopes, 
to find the lifetimes of the newly discovered super heavy 
elements \cite{superexp1} 
and sometimes even to understand the early cosmology of the universe 
\cite{cosmo}. Its many ramifications reached the realm of atomic
physics \cite{krausz} with ultracold atoms \cite{coldatoms}, 
as well as chemistry \cite{miyazaki, mcmahon}
and biology \cite{miyazaki, devault}. 
In spite of the fact that tunneling is now widely applied in 
many fields, the question of how much time does the particle require to 
tunnel and what connection does it have with the physically measured 
lifetimes of spontaneously decaying objects still remains unanswered
in the cases where we encounter a bound state before tunneling.

In the present work we attempt to answer the above question. To be specific, 
within the framework of semi-classical approximations, we derive expressions 
for the times spent by a tunneling particle in front of the barrier as 
well as within the barrier. Examining the known expressions for the 
lifetimes of decaying (metastable) states (or resonances), 
a connection is found between 
the so-called dwell times in tunneling and the lifetimes of metastable 
states. The dwell time is considered to be a measure of the average time 
spent by a particle in a given region of space. The concept was first 
introduced by Smith \cite{smith} in the context of quantum collisions 
and to derive a lifetime matrix for multichannel resonances. In the 
one-dimensional case, it was first introduced by B\"uttiker \cite{butik}. 
Indeed, earlier in 1966, Baz had proposed \cite{baz} the use of Larmor 
precession as a clock to measure the duration of quantum mechanical 
collision events and Rybachenko \cite{ryba} had applied the method to 
the simpler case of particles in one dimension. The work of B\"uttiker was 
an extension of these works. B\"uttiker and Landauer also defined 
\cite{butikland} a traversal or interaction time of transmitted 
particles in tunneling, which, as we shall see later also finds a new 
physical significance in the tunneling process. 
More recently, the dwell time formalism for the transition   
from a quasilevel to a continuum
of states was discussed in the context of electron and alpha particle 
tunneling by P. J. Price \cite{price}. Some other recent applications 
of dwell time can be found in \cite{mewinnuspri}. 

In what follows we shall present the formalism connecting the dwell times 
with the decay widths (and hence the half-lives) and conclude with the help 
of realistic examples that the dwell times of 
transmitted alpha particles in the region before entering the barrier 
correspond to the half-lives of radioactive nuclei decaying by alpha decay. 
Apart from the dwell time, 
there exist other tunneling time concepts in literature
and the subject in principle is replete with controversies \cite{winphysrep}. 
The controversy stems from definitions of phase times or 
the so-called ``group delay times"  
which can also predict superluminal tunneling velocities. 
The Hartman effect \cite{winphysrep} for 
example, states that the tunneling time becomes independent of the thickness 
of the barrier length for thick enough barriers and thus results in 
unbounded tunneling velocities. Such controversies, however, are not relevant 
for the calculations done in the present work.  
We do not evaluate the phase time but rather the dwell time and the 
barrier region itself 
makes a small contribution to the total dwell time 
(with the bulk coming from the region in front of the barrier). 
Hence we do not expect saturation effects such as the Hartman effect 
resulting in interpretations of superluminal alpha propagation. 

Though the nuclear potentials used in this work are not the most modern ones,
these potentials and the semiclassical approach \cite{book} 
used suffice to 
illustrate the main findings of this work. 
There exist potentials for which the semiclassical WKB approach 
may not be appropriate. 
However, it is known to work reasonably well for the alpha tunneling problem. 
This approach is indeed commonly used in literature for the evaluation of 
the standard Gamow factor and prediction of half-lives of heavy and
super heavy nuclei \cite{xuren}. We shall see below that the dwell
time (which one would generally not expect to be an experimentally 
measurable quantity) 
in the region in front of the barrier corresponds to the 
commonly used definition of the measured half-life. 
Since the WKB approximation 
is widely used for the calculation of half-lives, we too use the 
WKB wave function for the evaluation of dwell times in the present work. 

For an arbitrary barrier $V(x)$ in one-dimension (a framework
which is also suitable for spherically symmetric problems), 
confined to an interval 
$(x_1, x_2)$, the dwell time is given by the number of particles in 
the region divided by the incident flux $j$:
\begin{equation} 
\tau_D\,=\,{\int_{x_1}^{x_2}\,|\Psi(x)|^2\,dx \over j}\, .
\end{equation}
Here $\Psi(x)$ is the time independent solution of the Schr\"odinger equation 
in the given region. Though we shall restrict to using a semiclassical 
approximation for the wave function in the present work, one can always 
define a dwell time as above in a given region of space, be it with an exact 
or an approximate wave function. 
The standard definition of dwell time is the time spent in the region 
$(x_1, x_2)$ regardless of how the particle escaped (by reflection or 
transmission) and $j = \hbar \,k_0 \,/\mu$ \cite{footn}  
(with $k_0 = \sqrt{2 \mu E} / \hbar$) for a free particle. 
However, one can also define 
transmission and reflection dwell times for the particular cases when the 
particle is bound in a region and later either got transmitted or reflected. 
The flux $j$ in these cases would get replaced by the transmitted or 
reflected fluxes, 
$j_T = \hbar \, k_0 |T|^2/ \mu$ and $j_R = \hbar \, k_0 |R|^2 / \mu$ 
\cite{mario} respectively. The current $j_T$ is the particle's flux in the 
region III (see Fig. 1).
One would then obtain \cite{mario},  
\begin{equation}
{1 \over \tau_D} \, =\, {|T|^2 \over \tau_D} \, + \, {|R|^2 \over \tau_D} \, 
= \, {1 \over \tau_{D,T}} \, + \, {1 \over \tau_{D,R}}
\end{equation}
where $|T|^2$ and $|R|^2$ are the transmission and reflection coefficients 
(with $|T|^2 \, +\, |R|^2 \,=\, 1$ due to conservation of probability) and 
$\tau_{D,T} = \int |\Psi|^2 dx / j_T$ and 
$\tau_{D,R}  = \int |\Psi|^2 dx / j_R$,  
define the transmission and reflection dwell times 
respectively. The traversal time defined by B\"uttiker is somewhat different 
and is given as, 
\begin{equation}
\tau_{trav} (E)\,=\,\int_{x_1}^{x_2}\,\,{\mu \over \hbar \, k(x)}\,\,dx
\end{equation} 
where, $k(x)\,=\,{\sqrt{ 2\mu\,(|V(x)\,-\,E|)} / \hbar}$ with $E$ being 
the kinetic energy of the tunneling particle and $\mu$ the reduced mass. 
Having defined the tunneling times relevant to the present work, we shall now 
apply them to the study of the alpha decay of nuclei.

We study the alpha decay of nuclei as a tunneling of the $\alpha$ through
the potential barrier of the alpha-daughter nucleus system using a 
semi-classical approach. 
Typically, one considers the tunneling of the $\alpha$
through a spherically symmetric $r$-space potential (see Fig. 1) 
of the form,
$V(r)\, =\, V_n(r)\,+\,V_c(r)\,+\,{\hbar^2\,(l\,+\,1/2)^2 / 2 \mu\,r^2}$,
where $V_n(r)$ and $V_c(r)$ are the attractive nuclear and 
repulsive Coulomb parts of the
$\alpha$-(daughter) nucleus potential, $r$ the distance between
the centres of mass of the daughter nucleus and alpha and $\mu$
their reduced mass. The last term
represents the Langer modified centrifugal barrier \cite{langer}.  
\begin{figure}[ht]
\includegraphics[width=8cm,height=5cm]{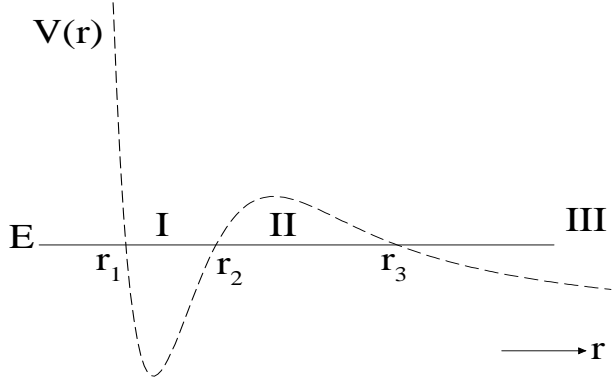}
\caption{
Typical potential $V(r)$ in an alpha-nucleus tunneling
problem. $r_1$, $r_2$ and $r_3$ are the classical turning points for a 
given kinetic energy $E$ of the tunneling particle.} 
\end{figure}
Writing the wave functions in region I and region II 
(up to a normalization factor) using the 
semi-classical Wentzel-Kramers-Brillouin (WKB) 
approximation \cite{landau}, 
\begin{eqnarray}
\Psi_{\rm I} (r) \,=\, {2\, \sqrt{k_0} 
\over \sqrt{k(r)}} \,\, {\rm cos} \, \biggl [ 
\int_{r}^{r_2} \, dr^{\prime} \, k(r^{\prime}) \, - \, {\pi \over 4} \, 
\biggr ] \\ \nonumber
\Psi_{\rm II} (r) \, =\, {\sqrt{k_0} \over \sqrt{\kappa (r)}} \,\, 
{\rm exp} \biggl [ \,-\, \int_{r_2}^r \, dr^{\prime} \, 
\kappa (r^{\prime}) \, \biggr ] \, ,   
\end{eqnarray}
where $k(r)\,=\,{\sqrt{ 2\mu\,(E\,-\,V(r))} / \hbar}$ and 
$\kappa(r)\,=\,{\sqrt{ 2\mu\,(V(r)\,-\,E)} / \hbar}$, 
the dwell times $\tau_D^{\rm I}$ and $\tau_D^{\rm II}$ 
in regions I and II respectively are given as follows: 
\begin{eqnarray}\label{dwell1}
\tau_D^{\rm I} (E) \, &=& \, {4 \, \mu \over \hbar } \, \int_{r_1}^{r_2} \, 
{dr \over k(r)} \, {\rm cos}^2 \biggl [ \, \int_r^{r_2} \, dr^{\prime} \, 
k(r^{\prime}) \, - \, {\pi \over 4} \, \biggr ]\\ \nonumber    
& \simeq &  \, {2 \, \mu \over \hbar } \, \int_{r_1}^{r_2} \, 
{dr \over k(r)} \, = \,2\, \tau_{trav}^{\rm I} (E) 
\end{eqnarray}
where the second line follows from replacing the squared cosine term 
by (1/2). The dwell time in region II is, 
\begin{equation}\label{dwell2}
\tau_D^{\rm II} (E) \, = \, 
{\mu \over \hbar}  
\, \int_{r_2}^{r_3} \, {dr \over \kappa(r)} \, {\rm exp} \,\biggl 
[ - 2 \, \int_{r_2}^r \, dr^{\prime} \, \kappa (r^{\prime}) \biggr ] \, .
\end{equation} 
Though the semiclassical WKB approach which is still often 
used for the evaluation of decay widths \cite{wkbrefs} is 
sufficient for the objectives of the present work, better methods such as
the Gamow-state formalism \cite{nazar} do exist. 
In fact, earlier, starting in the seventies, 
the general formula for the lifetime 
of a nucleus decaying by $\alpha$-decay was obtained on the basis of a
Gamow-state formalism in \cite{kadmen}. 
If we now consider the standard definition of the WKB decay width 
\cite{gurvitz},
\begin{equation}\label{gurwidth}
\Gamma (E)\,=\,P_{\alpha} \, \,
{\hbar^2 \over 2 \,\mu}\,\,\biggl[\,\int_{r_1}^{r_2}\,
{dr \over k(r)}\,\biggr]^{-1}\,e^{-2\int_{r_2}^{r_3}\, \kappa(r)\,dr} \, , 
\end{equation} 
($P_{\alpha}$ is the pre-formation probability of the alpha cluster inside the
radioactive nucleus decaying later by alpha particle emission) 
and compare it with equation (\ref{dwell1}), taken along with the fact that 
the transmission coefficient $|T|^2 = e^{-2\int_{r_2}^{r_3}\,\kappa(r)\,dr}$, 
we arrive at the first main result of this work, 
\begin{equation}\label{result1}
\Gamma (E) \, = \, P_{\alpha} \, 
\hbar \, |T|^2 \, \, [\tau_D^I (E)]^{-1} \, = \,P_{\alpha} \, \hbar \, \,
[\tau_{D,T}^I (E)]^{-1} \, .
\end{equation}
The decay width is given by the inverse of the {\it transmission dwell time} 
in the region in front of the barrier. This implies that the 
half-life (which is evaluated at the energy $E = Q$, where $Q$ is the 
amount of energy released in the decay), 
\begin{equation}\label{result2} 
\tau_{1/2} \, =\, {\hbar \, {\rm ln \,2} \over \Gamma } \, = \, 
{{\rm ln \, 2} \over P_{\alpha}}\,\, \, \tau_{D,T}^I \, ,  
\end{equation}
is essentially given by the {\it transmission dwell time} in region I. 
It is of further interest to note that the frequency of assaults at the 
barrier, $\nu$, can be written as the inverse of the time required
to traverse the distance back and forth between the turning points $r_1$ and
$r_2$ as \cite{froeman},
\begin{equation}\label{period}
\nu\,=\,{\hbar \over 2\,\mu}\,\biggl[\,\int_{r_1}^{r_2}\, 
{dr \over k(r)}\,\biggr]^{-1}\,.
\end{equation}
However, from equation 
(\ref{dwell1}), it follows that $\nu\,=\, [\tau_D^I]^{-1}$ 
and taken along with equation (\ref{result1}) with $P_{\alpha} = 1$, 
we can see that the number of 
assaults that the particle makes before tunneling is
$N_a \,=\, \nu \, \tau_{D,T}^I \, = \, (|T|^2 )^{-1}$. 

In Table I, we list the transmission dwell times for 
four medium heavy nuclei with $l=0$ and
two recently studied super heavy nuclei \cite{superexp3}. 
In the absence of much information on the $l$ values of the super heavy 
nuclei studied, we assume the angular momentum $l = 0$. 
The case of the light nucleus $^8$Be (with $l =2$) 
is presented separately in Table II.
The alpha-alpha 
potential for $^8$Be is described analytically and 
is taken from \cite{satchler}. The alpha-nucleus potential for the medium 
and super heavy nuclei is constructed using a double folding model with 
realistic nucleon-nucleon ($NN$)interactions as given 
in \cite{satchler} and also used in some recent works \cite{m3yusers,wkbrefs}. 
An accurate estimate would however demand the latest available $NN$ potentials
\cite{goodnn} and considerations of the four-particle correlations 
in nuclei \cite{fourp}.  
The Coulomb potential is also obtained via a double folding procedure where the 
matter densities of nuclei are replaced by their charge densities.
The details of the potentials used in the present work 
can be found in \cite{we2}.  

We also list the transmission dwell times in region II,   
in the tables. 
To evaluate the total time spent in regions I and II, one would 
start with the definition of the dwell time as, 
\begin{equation}\label{taunew}
\tau_D^{\rm full} (E) \, = \,\int_{r_1}^{r_3}\, 
{|\Psi(r)|^2\,dr \over j}\, . 
\end{equation}
The lifetime of the decaying nucleus 
should in principle be related to the total 
transmission dwell time spent by the $\alpha$ in the two regions
and not just in region I as given by (\ref{result2}). 
Defining such a time as, say, 
\begin{equation}\label{sumtau} 
\tau_{D,T}^{full} (E) = 
\tau_{D,T}^{\rm I} (E) \, +\, \tau_{D,T}^{\rm II} (E)\, ,
\end{equation}
the half-life expression in (\ref{result2}) would rather take the form,
\begin{equation}\label{result3} 
\tau_{1/2}^{\rm full} \,= \, 
{{\rm ln \, 2} \over P_{\alpha}^{\rm full}}\,\, \, \tau_{D,T}^{\rm full} \, , 
\end{equation}
with $P_{\alpha}^{\rm full}$ being the preformation factor in this case.   
For the medium heavy nuclei, agreement with experimental half-lives is 
obtained with $P_{\alpha}^{\rm full} \sim 0.4 - 0.55$. 
These values are close to those 
obtained in literature \cite{xuren} with similar potentials. 
The $P_{\alpha}^{\rm full}$ values for the 
super heavy ones are around $0.2$, which could change with 
possible higher angular momenta for these nuclei. 
In principle, the values of $P_{\alpha}^{\rm full}$ depend on the details 
of the nuclear potentials which in turn affect the tunneling probabilities. 
The values of $P_{\alpha}^{\rm full}$ found here should hence 
not be interpreted as a constraint on
four-particle correlations \cite{fourp} 
in microscopic nuclear structure models.
A detailed study of the preformation factors for super heavy alpha emitters 
can be found in \cite{mohr}.  
The half-lives are evaluated 
at the experimental $Q$ values, i.e., at $E =Q$ in equation (\ref{result2}). 
\begin{table}[h]
\caption{\label{tab1} Comparison of the calculated half lives 
(from the transmission dwell times) with experiment, 
for medium and super heavy alpha emitters.}
\begin{tabular}{|l|l|l|l|l|l|l|l|}
\hline
Radioactive& Q value &ln2\,$\tau_{D,T}^{\rm I}$ 
&ln2\,$\tau_{D,T}^{\rm II}$ & ln2 \,($\tau_{D,T}^{\rm I}$ + 
$\tau_{D,T}^{\rm II} $) & $\tau_{1/2}$ (expt) & $P_{\alpha}^{\rm full}$ 
\footnotemark[1] 
& Number of \\ 
Nucleus & (MeV) &\,\,\,(s) & \,\,\,(s)  & \,\,\,\,(s) & \,\,\,\,(s) & 
& assaults\\ 
\hline
\hline
$^{108}_{52}$Te & 3.445 & 1.506  & 0.261  & 1.767  & 4.286  & 0.41 
& 7 $\times$ 10$^{21}$ \\ \hline
$^{169}_{77}$Ir & 6.151 & 0.466  & 0.069  & 0.535  & 1.28  & 0.42 & 
2 $\times$ 10$^{21}$ \\ \hline
$^{173}_{79}$Au & 6.836 & 9.29 $\times$ 10$^{-3}$  & 1.38 $\times$ 10$^{-3}$ & 
10.67 $\times$ 10$^{-3}$  & 26.6 $\times$ 10$^{-3}$  & 0.40 & 4 $\times$
 10$^{19}$ \\ \hline
$^{180}_{74}$W & 2.508 & 2.69 $\times$ 10$^{25}$   & 3.74 $\times$ 10$^{24}$ & 
3.06 $\times$ 10$^{25}$  & 5.68 $\times$ 10$^{25}$ & 0.54 & 1 $\times$ 
10$^{47}$ \\ \hline
\hline
$^{273}_{110}$Ds & 11.368 &  2.038 $\times$ 10$^{-5}$  & 
2.76 $\times$ 10$^{-6}$ &  2.32 $\times$ 10$^{-5}$  & 17 $\times$ 10$^{-5}$ 
& 0.14  & 9 $\times$ 10$^{16}$ \\ \hline
$^{277}$112 & 11.3 & 1.16 $\times$ 10$^{-4}$  & 1.52 $\times$ 10$^{-5}$ 
& 1.31 $\times$ 10$^{-4}$  & 6.90 $\times$ 10$^{-4}$ & 0.19  & 5 $\times$ 
10$^{17}$  \\ \hline
\end{tabular}
\footnotetext[1]{\,\,ln2\,($\tau_{D,T}^{\rm I}$ + $\tau_{D,T}^{\rm II}$) /
$\tau_{1/2}{\rm (expt)} = P_{\alpha}^{\rm full}$}
\end{table}
\noindent
\begin{table}[h]
\caption{\label{tab2} Same as Table I but for the light nucleus $^8$Be.}
\begin{tabular}{|l|l|l|l|l|l|l|l|}
\hline
Radioactive& Q value &ln2\,$\tau_{D,T}^{\rm I}$ 
&ln2\,$\tau_{D,T}^{\rm II}$ & ln2 \,($\tau_{D,T}^{\rm I}$ + 
$\tau_{D,T}^{\rm II} $) & $\tau_{1/2}$ (expt) & $P_{\alpha}^{\rm full}$ 
\footnotemark[1] 
& Number of \\ 
Nucleus & (MeV) &\,\,\,(s) & \,\,\,(s)  & \,\,\,\,(s) & \,\,\,\,(s) & 
& assaults\\ 
\hline
\hline
$^{8}$Be (l=2) & 3.1218 & 3.7 $\times$ 10$^{-22}$ & 2.56 $\times$ 10$^{-22}$ & 
6.26 $\times$ 10$^{-22}$  & 3.02 $\times$ 10$^{-22}$ & 2.1 & 1 \\ \hline
         &  &  &   &  &  &  & \\ \hline
& &ln2\,$\tau_{D}^{\rm I}$ 
&ln2\,$\tau_{D}^{\rm II}$ & ln2 \,($\tau_{D}^{\rm I}$ + 
$\tau_{D}^{\rm II} $) & &  
&  \\ 
 & &\,\,\,(s) & \,\,\,(s)  & \,\,\,\,(s) &  & 
& \\  \hline
 &  & 1.59 $\times$ 10$^{-22}$ & 2.19 $\times$ 10$^{-22}$ & 3.78 $\times$ 
10$^{-22}$ & 3.02 $\times$ 10$^{-22}$ & 1.25 & 1 \\ \hline
\end{tabular}
\footnotetext[1]{\,\,ln2\,($\tau_{D,T}^{\rm I}$ + $\tau_{D,T}^{\rm II}$) /
$\tau_{1/2}{\rm (expt)} = P_{\alpha}^{\rm full}$}
\end{table}

We find that the major bulk of the half-life of a 
medium or super heavy radioactive nucleus is spent in region I, in front of
the barrier before tunneling. Though the time spent inside 
the barrier is much smaller, it is not negligible and should be added to 
that in region I to obtain the total time spent in tunneling.   
The number of knocks (assaults) made at the 
barrier is inversely proportional to the transmission coefficient which 
for the heavy nuclei is extremely small leading to a huge number of assaults 
by the alpha at the barrier. With $|R|^2$ being almost unity in these cases, 
the reflection dwell times, $\tau_{D,R}$ 
as well as the average dwell times $\tau_{D}$ are orders of magnitude smaller
as compared to $\tau_{D,T}$ as well as the half-lives.  
The times spent in region II (the barrier) 
are always an order of magnitude smaller. The case of $^8$Be 
(with $l=2$ studied here) is however, very different. 
With the transmission coefficient being reasonably big 
($|T|^2 = 0.86$ at the $Q$ value of $3.1218$ MeV), the times, 
$\tau_{D,R}$, $\tau_{D,T}$ and $\tau_D$ are all comparable and of the same 
order of magnitude in both regions. 
It makes more sense then,  
to compare the standard average dwell time definition with the 
experimental half-life.
An interesting implication of the large transmission coefficient here is 
that the alpha in $^8$Be escapes after one knock. 

In conclusion, we can say that the present work provides a new look at the
physics of the quantum tunneling problem in general and the alpha decay 
problem in particular.
In principle, one does not really know how alpha particles or 
clusters of heavier nuclei are formed inside the nucleus.
In fact, there are arguments that they can be formed in virtual states 
\cite{zelev1} different from the states of those objects in free space.
Such objects could undergo some
restructuring in the nucleus \cite{zelev2} which 
can then not be described by a simple potential consideration. 
The present work does not attempt to address these issues. 
The conclusions drawn in this work are hence limited to the picture of 
a preformed cluster (say the alpha) inside the nucleus, 
which tries to tunnel through a potential barrier. 
Further, we note that in contrast to the 
lifetime definition for the decay of 
an elementary particle, 
say muon decay, the definition of lifetime in the context of a tunneling 
problem is somewhat different. Here the tunneling object has to traverse a 
region of space, thus giving rise to the question of how long does it 
require for a particle to tunnel 
(for a general discussion, see \cite{razavy}). 
The dwell time concept answers this question. In the alpha tunneling 
problem studied in the present work, it is found that the dwell time 
of the $\alpha$ particle in front of the barrier as well as within the 
barrier is important.  
This can be relevant not only 
to the alpha radioactive nuclei of medium mass and 
the recent discoveries of super heavy nuclei, but also to 
other branches of science and physics where the quantum concepts of 
time in conjunction with bound states and 
tunneling might prove fruitful to have a measure of the speed of a reaction.


\begin{thebibliography}{99}
\bibitem{footn}
We do not explicitly mention the normalization factor of the wave function 
and its corresponding appearance in the current density since we are 
only interested in evaluating dwell times, where the normalization cancels 
in the ratio.
\end{thebibliography}
\end{document}